\documentclass[12pt]{article}
\def\simge{\mathrel{%
   \rlap{\raise 0.511ex \hbox{$>$}}{\lower 0.511ex \hbox{$\sim$}}}}
\def\simle{\mathrel{
   \rlap{\raise 0.511ex \hbox{$<$}}{\lower 0.511ex \hbox{$\sim$}}}}

\newcommand{\nc}{\newcommand}
\nc{\beq}{\begin{equation}}
\nc{\eeq}{\end{equation}}
\nc{\beqa}{\begin{eqnarray}}
\nc{\eeqa}{\end{eqnarray}}

\begin{document}
\setlength{\baselineskip}{24pt}

\title{\large {\bf The 2+1 Dimensional NJL Model at Finite Temperature }}

\author{
Thomas Appelquist\thanks{thomas.appelquist@yale.edu}\\
{\small Department of Physics,
Yale University, New Haven, CT 06511.} \\ \\
Myck Schwetz\thanks{ms@bu.edu}
 \\ {\small Department of Physics,
Boston University, Boston, MA 02215.} \\ \\
}

\maketitle

\begin{picture}(0,0)(0,0)
\put(360,344){YALE-P8-00}
\put(360,322){BUHEP-00-16}
\end{picture}
\vspace{-24pt}

\begin{abstract}

 We decribe properties of $2+1$-dimensional Nambu-Jona-Lasinio (NJL)
models at finite-temperature, beginning with the model with a
discrete chiral symmetry. We then consider the model with a
continuous $U(1) \times U(1)$ chiral symmetry, describing the
restoration of the symmetry at finite temperature. In each case,
we compute the free energy and comment on a recently proposed
constraint based upon it. We conclude with a brief discussion of
NJL models with larger chiral symmetries.

\end{abstract}

\section{Introduction}

 The behavior of QCD at finite
temperature and density has received renewed attention 
recently with the coming of the RHIC experimental program. 
The study of this phase structure is difficult for any 
strongly coupled quantum field theory, and it can be helpful 
to examine the problem in nontrivial but tractable models. 
In particlular, theories in less than four space-time dimensions can offer
interesting and complex behavior as well as tractability, often in the
form of large-N expansions. In the case of three space-time dimensions,
they can even  be directly physical, describing various planar condensed
matter systems.
Models of interest for the study of finite temperature 
{\it and} density involve fermionic degrees of freedom and fall into
two broad classes. One utilizes four-fermion
interactions of the Nambu-Jona-Lasinio (NJL), or Gross-Neveu,
form, and the other includes three dimensional gauge theories and
closely related Thirring models.

 We restrict our attention in this paper to NJL models at finite
temperature $T$. The features we discuss are interesting in their
own right and will play an important role in understanding
temperature-density phase diagrams for these models. All the
analysis will make use of a $1/N_f$ expansion where $N_f$ is the
number of fermion species. The theories under consideration are
renormalizable in the $1/N_f$ expansion unlike in the loop
expansion \cite{RWPreview}. The expansion is
reliable for all temperatures $T$ except those in the vicinity of 
a certain critical temperature  $T_c$.

The question of interest is the
spontaneous breaking of chiral symmetry. Although this is in some
ways easier to handle in three space-time dimensions ($3D$) than
in four, there is an important subtlety in the case of {\it
continuous} symmetry. The Coleman-Mermin-Wagner (CMW) theorem
stipulates that the spontaneous breaking of a continuous symmetry
can not happen in $3D$ at finite $T$ \cite{CMW}. This statement is
related to the fact that it is impossible to write down a
consistent theory of massless scalars in $2D$. If a spontaneous
breaking of continuous symmetry were to happen at finite $T$, then
one would be faced with this problem at momentum scales below $T$,
i.e. it would be impossible to construct an effective $2D$ theory
of the Goldstone bosons' zero-modes.

  We first review the behavior of the $3D$ NJL model with a
discrete chiral symmetry, where the CMW problem does not arise.
The broken symmetry at $T = 0$  remains broken at finite $T$ up to
a critical value $T_c$. We then turn to the $3D$ NJL model with a
continuous $U(1)\times U(1)$ chiral symmetry, exploring symmetry
restoration at all $T >0$. A critical temperature $T_c$ 
continuous to exist, but it now marks a
transition from the ordinary symmetric phase at high temperature to a low
temperature Kosterlitz-Thouless phase that is also chirally symmetric. The
broken phase exists only at $T=0$.

For each model, we also compute the thermodynamic free energy and 
enumerate the thermodynamic degrees of freedom. In a recent paper
\cite{ACS}, the free energy $F(T)$ was used as the basis for a proposed
constraint on the behavior of asymptotically free theories. It was
observed that for any theory governed by a fixed point in the
ultraviolet (UV) or infrared (IR), the dimensionless quantity
$f(T) \equiv -2\pi [F(T)-F(0)] / \xi(3)T^3$ approaches a finite
value in the corresponding limit 
( $f_{UV} = f(T \rightarrow \infty)$ and $f_{IR} = f(T \rightarrow 0)$ ), 
and counts the (effectively massless) 
degrees of freedom if the fixed point is trivial. It
was noted that $f_{UV} \geq f_{IR}$ for all known asymptotically
free theories in which the IR behavior is also free, or weakly
interacting, allowing $f_{IR}$ to be computed. It was conjectured
that this is true for all asymptotically free theories. 
The theories considered in this paper are 
governed by nontrivial UV fixed points and apriori may or may not 
satisfy this condition.

The outline of the paper is as follows: 
in Section 2, we review the NJL model with a discrete chiral
symmetry; in Section 3, we treat the NJL model with a continuous
$U(1)\times U(1)$ chiral symmetry; we summarize and briefly
describe NJL models with larger chiral symmetries in Section 4.

\section{ Discrete Chiral Symmetry}

The NJL model with a discrete chiral symmetry for $N_f$ copies of
Dirac fermions is described by the following
Lagrangian\footnote{We adopt the notation of \cite{RWPreview} for
$\gamma$-matrices: $\gamma^{\mu} = \sigma^{\mu} \otimes
\left(\begin{array}{cc}
I & 0 \\
0 & -I
\end{array} \right)$ and $i \gamma_5 =
\left(\begin{array}{cc}
0 & I \\
-I& 0
\end{array} \right)$
in $(2+1)$-dimensional Minkowski space with $(1, -1, -1)$
signature.}: \beq \label{njldis} {\cal L}~=~ i \,
\bar{\Psi}\partial\!\!\!/ \Psi ~+~ {g_0^2 \over 2
N_f}\,\Big(\bar{\Psi} \Psi\Big)^2, \eeq where $\Psi$ may be taken
to be a $4 N_f$-component fermion field. The discrete symmetry is
$\Psi \rightarrow \gamma_5 \Psi$, and $g_0$ is the coupling. We
analyze the model using the $1/N_{f}$ expansion.

We work first to leading order and then note that for this
discrete-symmetry model the qualitative behavior is not modified
in higher orders. It is convenient to introduce an auxiliary field
$\sigma$ coupled to $\bar{\Psi}\Psi$. The finite temperature
effective potential as a function of $\sigma$ may be computed to
this order by integrating out the fermions: \beq \label{Vdis}
V_{eff}(\sigma) /N_f ~=~ {\sigma^2 \over 2 g_0^2}~+~ i\, Tr ln
\Big(i\,\partial\!\!\!/ - \sigma), \eeq with the second term
evaluated at finite temperature.

The only cutoff dependence is in the renormalization of $g_0$, and
it may be carried out at zero temperature. Extremizing the
zero-temperature effective potential ($\partial V_{eff}/\partial
\sigma = 0$) gives \beq \label{gapdis} {1 \over g_0^2} ~=~ 4
\int^\Lambda {d^3p \over (2\pi)^3}\, {1 \over p^2 +
\sigma_0^2}~=~{ 2\over \pi^2} \Big(\Lambda - \sigma_{0}
tan^{-1}{\Lambda\over\sigma_0}\Big) \eeq where $\Lambda$ is an
ultraviolet cutoff and where the extremal value $\sigma_0$
describes the fermion mass gap at zero temperature. As $g_{0}^{2}
\Lambda$ approaches the critical value $\pi^2 / 2$ from above,
$\sigma_0 / \Lambda \rightarrow 0 $. Equivalently, the cutoff may
be removed holding $\sigma_0$ fixed providing that  $g_{0}^{2}
\Lambda$ is tuned to $\pi^2 / 2$. With this prescription, it can
be seen that the high energy behavior of the theory is described
by a nontrivial but weak ultraviolet fixed point. That is, a
finite effective four-fermion interaction $\bar{g}^{2}(p)/2 N_f$
is induced such that for $p >> \sigma_0$, the dimensionless
quantity $\bar{g}^{2}(p)\cdot p \rightarrow {\cal O}(1)$.

To leading order in $1/N_{f}$, the
 zero-temperature effective
potential is now
 \beq \label{VdisT0} V_{eff}^{T=0}(\sigma)/N_f
~=~ {1 \over 3 \pi}\,\sigma^3 ~-~ {\sigma_0\over 2 \pi}\,
\sigma^2~, \eeq which has a minimum at $\sigma = \sigma_0$ and is
convex for $\sigma > \sigma_0/2$. At finite temperature, the
leading order effective potential takes the form: \beq
\label{VdisT} V_{eff}(\sigma)/N_f ~=~ {1 \over 3 \pi}\,\sigma^3
~-~ {\sigma_0\over 2 \pi}\, \sigma^2 ~-~ T^3 \, \int_0^\infty
{dx\over \pi}\, ln\Big(1~+~exp[-\sqrt{x + {\sigma^2 \over
T^2}}]\Big)~. \eeq It may be extremized to give the temperature
dependent gap equation \beq \label{gapdisT} \sigma_{T} ~=~
\sigma_0 ~-~ 2 T \, ln \Big(1 ~+~exp[-{\sigma_{T} \over T}]\Big)
~=~\sigma_0 ~-~ 2 T \sum_{k=1}^{\infty} {(-1)^{k+1} \over k}\,
exp[-k {\sigma_{T} \over T}] ~. \eeq This leads to the critical
temperature $T_c=\sigma_0/(2\, ln2)$, above which $\sigma_{T}$
vanishes. The existence of a finite $T_c$ for a discrete symmetry
in $2+1$ dimensions is perfectly acceptable and this qualitative
feature is not changed by the higher order terms in the $1/N_f$
expansion.

Before discussing higher order terms, we comment on the thermodynamic
free energy of this model which 
is given by $F(T) = V_{eff}(\sigma_{T})$. Then,
to leading order in $1/N_f$, the $f(T)$ defined in the introduction will
take the form
\beqa \label{fdisT}
f(T)~=~-\, {2 N_f \over \zeta(3) T^3} \, \Big(\, {\sigma_{T}^3
\over 3} - {\sigma_0 \over 2}\, \sigma_{T}^2 + {\sigma_0^3 \over
6} ~+~ 2 T^3 \sum_{k=1}^{\infty} {(-1)^{k+1} \over k^3}\, \Big(1 +
k {\sigma_{T} \over T}\Big)\, exp[-k {\sigma_{T} \over T}]\,
\Big)~, \eeqa where $\sigma_{T}$ is determined by the gap equation
(\ref{gapdisT}). It can be shown that this function increases
monotonically from $f_{IR} = 0$ at $T = 0$ to $f_{UV} = 3 N_f $ at
$T = \infty$. The infrared value, $f_{IR} = 0$, is expected
because a mass gap develops for for $T < T_c$. The ultraviolet
value, $f_{UV} = 3 N_f$, counts the number of fermionic degrees of
freedom ($4~N_f$) times the Fermi-Dirac factor (3/4). This is
expected because the model is governed in the ultraviolet by a
weak UV fixed point -- there will be higher order corrections in
the $1/N_f$ expansion. Monotonicity may be established by noting
that \beqa \label{mon} {d f(\sigma_{T}, T)\over d T} ~=~ {\partial
f \over \partial \sigma_{T}}\, {\partial \sigma_T \over  \partial
T} ~+~  {\partial f \over \partial
T}~=~~~~~~~~~~~~~~~~~~~~~~~~~~~~~~~~~~~~~~~~~~~~~~~~
  \nonumber  \\
~=~ {1 \over \zeta(3) T^4}\,
\Big( \sigma_{T}^3 - {3 \sigma_0 \over 2} \sigma_{T}^2 + {\sigma_0^3 \over
2}\Big)
~+~ {2 \over \zeta(3)}\, {\sigma_{T}^2 \over T^3} \, ln\Big(1 +
exp[-{\sigma_{T} \over T}]\Big)~.
\eeqa
The first term is always non-negative since $\sigma_{T} \leq
\sigma_0$; the second term is always positive.

 Finally we describe the effect of the higher order terms
in the $1/N_f$ expansion. Through next order, the effective potential
is given by \beq \label{VdisNO} V_{eff}(\sigma)/N_f ~=~ {\sigma^2 \over 2
g_0^2}~+~ i\, Tr ln \Big(i\,\partial\!\!\!/ - \sigma\Big) ~+~ {i
\over 2N_f} Tr~ln~ D_{\sigma}^{-1}, \eeq where
$D_{\sigma}^{-1}(p^2)$ is the inverse $\sigma$ propagator computed
in a background $\sigma$ field at finite temperature $T$.
 At zero
temperature, this propagator is given by \beqa \label{sigampropL}
D_{\sigma}^{-1}(p^2)~=~ N_{f} \, \Big(\, {1 \over g_0^2} ~-~i \int
{d^3q\over (2\pi)^3} Tr {1 \over q\!\!\!/ - \sigma} {1 \over
q\!\!\!/ - p\!\!\!/  - \sigma} \,\Big), \eeqa Using the leading
order gap equation (\ref{gapdis}) and taking the limit $\Lambda
\rightarrow \infty$ with $\sigma_0$ fixed, $D_{\sigma}^{-1}(p^2)$
takes the form \beqa \label{sigmaprop} D_{\sigma}^{-1}(p^2)~=~
{N_{f}\over \pi} \, \Big[ \,\sigma - \sigma_0 ~+~ {p^2 + 4
\sigma^2 \over 2 \sqrt{p^2}} \, tan^{-1} {\sqrt{p^2} \over 2
\sigma}\, \Big]~. \eeqa 
%Note that $D_{\sigma}^{-1}(p^2)$ is
%positive definite for all Euclidean momenta if $\sigma >
%\sigma_0/2$, i.e. if $\sigma$ is above the inflection point of Eq.
%\ref{VdisT0}. 
For nonzero temperature $D_{\sigma}^{-1}(p^2)$ is
given in Ref. \cite{MPG}. 
%It, too, is positive definite if $\sigma > \sigma_0/2$. 
Expressions, similar to Eq. \ref{VdisNO} may be written down for higher
order terms.

For $T$ well above $T_c$, the $1/N_f$ expansion for $V_{eff}(\sigma)$ 
may be seen to converge
for large $N_f$. As $T \rightarrow T_c$ from above, however, $m^2
\sim T - T_{c}$. In this limit, higher order terms in
$V_{eff}(\sigma, \pi)$ become singular and trigger the breakdown
of the $1/N_f$ expansion. They may be described by an effective
2D, Landau-Ginzburg theory, consisting of the zero mode of the
$\sigma$ field, relevant at scales below $T$. 
The terms in the Landau-Ginzburg Lagrangian, in
addition to the common mass term $m^2 \sigma^2/2 $, are a kinetic
term, and a $\lambda \sigma^4/4! $ interaction term. Each may be
computed to any order in $1/N_f$ by integrating out the fermions.
To leading order, each arises with coefficient $N_f$.

Using this effective 2D theory and counting infrared powers, it
may be seen that the effective dimensionless expansion parameter
for $V_{eff}(\sigma)$, for small $\sigma$, is of the form
 \beq
  \label{IRexp}
  {{1}\over{N_f}}{{T^2}\over{M^2}},
\eeq where $M^2 = m^2 + \lambda \sigma^{2}$, up to logarithmic
corrections. Thus in the neighborhood of the origin, the $1/N_f$
expansion breaks down for $|T - T_{c}| / T  \simle 1/N_f$. In
particular, it breaks down for the free energy and $f(T)$. This is
true also as $T$ approaches $T_c$ from below, as indicated by the
absolute value sign \cite{Sachdevbook}.

The $1/N_f$ expansion remains convergent as long as $|T_c - T| / T
>> 1/N_f$. In the high temperature limit, it correctly describes
the effective potential and free energy in the symmetric phase,
leading to small corrections to the leading order estimate
$f_{UV}= 3N_f$. In the low temperature limit, it describes the
broken phase, and because of the mass gap, leads to $f_{IR} = 0$
to all orders. The inequality $f_{UV} \geq f_{IR}$ is satisfied.

\section{ Continuous $U(1)\times U(1)$ Chiral Symmetry}

The NJL model with a continuous $U(1)\times U(1)$ symmetry $\Psi
\rightarrow exp(i\alpha \gamma_5) \Psi$, $\Psi \rightarrow
exp(i\beta)\Psi$ is described by the Lagrangian \cite{RWPreview}:
\beq \label{njlcon} {\cal L}~=~ i \, \bar{\Psi}\partial\!\!\!/
\Psi ~+~ {g_0^2 \over 2 N_f}\,\Big[\,\Big(\bar{\Psi} \Psi\Big)^2
~-~ \Big(\bar{\Psi} \gamma_5 \Psi\Big)^2\, \Big]. \eeq The model
may be analyzed to leading order in the $1/N_f$ expansion by
introducing auxilliary fields $\sigma$ and $\pi$ and integrating
out the fermions. The leading-order, finite-temperature effective
potential takes the form \beq \label{Vcon} V_{eff}(\sigma,
\pi)/N_f ~=~ {\sigma^2 + \pi^2 \over 2 g_0^2}~+~ i\,
Tr~ln~\Big(i\,\partial\!\!\!/ - \sigma - i \gamma_5 \pi\Big) \eeq

As in the case of discrete symmetry, the renormalization may be
performed by extremizing the effective potential ($\partial
V_{eff}/\partial \sigma = 0$ and $\partial V_{eff}/\partial \pi =
0$) at zero temperature. The vacuum expectation values may be
rotated using the chiral symmetry so that $\pi$ has a vanishing
expectation value. The zero temperature gap equation then takes
the same form as in the discrete-symmetry model: 
\beq
\label{gapcon} 1 ~=~ { 2\over \pi^2}\,  \Big(g_0^2 \Lambda\Big) \,
\Big(1 - {\sigma_0 \over \Lambda}
tan^{-1}{\Lambda\over\sigma_0}\Big), \eeq where $\Lambda$ is an
ultraviolet cutoff. 
As before, $\sigma_0$ may be held fixed in the ``continuum''
limit $\Lambda \rightarrow \infty$ ($g_{0}^{2}
\Lambda \rightarrow \pi^2 / 2$) describing the fermion mass
gap at zero temperature. The chiral symmetry is broken and the
$\pi$ field describes the associated Goldstone boson. The zero
temperature effective potential to leading order in $1/N_f$ now
reads: \beq \label{leadinggap} V_{eff}(\sigma, \pi)/N_f ~=~ {1
\over 3 \pi}\,\Big(\sigma^2 + \pi^2 \Big)^{3/2} ~-~ {\sigma_0\over
2 \pi}\, \Big(\sigma^2 + \pi^2 \Big)~, \eeq which has 
degenerate minima at $\sigma^2 + \pi^2  = \sigma_0^2$ and is 
convex as a function of two variables if 
$\sigma^2 + \pi^2  \geq \sigma_0^2$.

%implying 
%that (\ref{leadinggap}) is unphysical for 
%$\sigma^2 + \pi^2 < \sigma_0^2$.

At finite temperature, the leading order effective potential has
the same form as Eq. (\ref{VdisT}), with the replacement
$\sigma^2 \rightarrow \sigma^2 + \pi^2$: 
\beq
 \label{VcontT}
  V_{eff}(\sigma, \pi)/N_f ~=~ {1 \over 3
\pi}\,(\sigma^{2} + \pi^{2})^{3/2} ~-~ {\sigma_0\over 2 \pi}\,
(\sigma^{2} + \pi^{2}) ~-~ T^3 \, \int_0^\infty {dx\over \pi}\,
ln\Big(1~+~exp[-\sqrt{x + {(\sigma^{2} + \pi^{2}) \over
T^2}}]\Big)~.
 \eeq
Extremizing it suggests that, as in the case of discrete symmetry,
the broken symmetry of the zero temperature theory remains broken
at finite temperatures below a critical value $T_c =
\sigma_{0}/2~ln2$. But unlike the discrete case, this conclusion
cannot be correct since the zero-mode of the associated Goldstone
boson would describe an effective 2D theory with a spontaneously
broken continuous symmetry -- in contradiction with the
Coleman-Mermin-Wagner theorem.

We explore the resolution of this problem by first noting that as
in the case of discrete symmetry, the $1/N_f$ expansion is
convergent as long as $T$ is not near the transition temperature
$T_c$  ( $|T_c - T| / T >> 1/N_f$ ). The next order term in
$V_{eff}(\sigma, \pi)/N_f$, for example, may be written in the
form 
\beq 
\label{1/N}
{i \over 2N_f} Tr~ln~D_{\sigma}^{-1} + {i
\over 2N_f} Tr~ln~D_{\pi}^{-1},
 \eeq
where $D_{\sigma}^{-1}$ and $D_{\pi}^{-1}$ are functions of $T$,
$T_c$, momentum, and field strength \cite{MPG}. This term can be
seen to be of order $1/N_f$ for $|T_c - T| / T >> 1/N_f$. But as
in the case of discrete symmetry, the $1/N_f$ expansion breaks
down due to infrared singularities when 
$|T - T_{c}| / T \simle 1/N_f$, 
with the singular terms describable by an effective
2D Landau-Ginzburg Lagrangian.

For $T >> T_c$, the theory is in the symmetric phase with a
convergent $1/N_f$ expansion. For $T << T_c$, even though the
$1/N_f$ expansion is again convergent, the continuous symmetry
model is, unlike the discrete symmetry model, not in the broken
phase. The chiral symmetry remains unbroken, although in a way
different than in the high temperature range. To see this, it is
convenient to use the following parametrization\footnote{ 
This transformation of variables must be handled carefully.
The jacobian of the transformation in the functional integral
leads to a cubically divergent term in the effective 
lagrangian of the form $\delta^3(0)\, ln \, \rho$, 
which serves only to cancel other such terms arising
in the quantum computation \cite{SW}. Without this 
cancellation, the jacobian term might be assumed \cite{YI}
to change the cutoff dependence  from that 
described in Eq. \ref{gapcon}, eliminating the existence of the 
continuum limit as described there.} 
of the auxiliary
fields for $T < T_c$~\cite{W}: 
\beq \label{param} 
\sigma ~+~ i\, \pi ~=~ \rho\, e^{i \theta}~. 
\eeq 
The leading order potential, Eq.
\ref{VcontT}, then depends only on $\rho$, and extremization leads
to a non-zero VEV $\rho_T$, equal to $\rho_0 \equiv \sigma_0$ 
at $T = 0$ and vanishing like $(T_{c}- T)^{1/2}$  
as  $T \rightarrow T_c$. The
fermion has mass $\rho_T$, the fluctuations of $\rho$ have mass $2
\rho_T$ and there is a massless scalar field $\theta$. 
%\footnote{
%The transformation (\ref{param}) produces a jacobian factor in the
%functional integral measure 
%\beqa 
%\label{measure} 
%{\cal D} \sigma(x) \, {\cal D} \pi(x) ~=~ {\cal D}\rho (x) \, {\cal
%D}\theta (x) \, exp\Big[\delta^3(0) \int d^3x \, ln \rho(x)\, 
%\Big]~, \nonumber 
%\eeqa
%but it contributes to the gap equation for $\rho$ only at the next
%to leading order in the $1/N_f$ expansion.}

This behavior implies symmetry breaking, however,
only if $\theta$ takes on
some fixed and non-zero vacuum value. Now as is well known, this
doesn't happen in 2D since quantum effects generate
logarithmically infrared divergent fluctuations. In the present
model at finite T , the same is true since the long distance
behavior is effectively 2D. To explore this in detail, we first restrict
attention to $T$ sufficiently below  $T_c$ ($(T_c - T) / T
>> 1/N_f$)  so that the $1/N_f$ expansion is a reliable tool .
 The realization of the symmetry may be studied by
examining the theory in the infrared, at momentum scales well
below $\rho_T$, where the fermions and the $\rho$ fluctuations may
be integrated out.  The only massless degree of freedom is
$\theta$, and if $\rho_T$ and $T$ are of the same order, the
non-zero Matsubara frequencies of $\theta$ may also be integrated
out. The effective low energy theory describing physics at momentum scales 
well below $T$ and $\rho_T$ is then a 2D
chiral Lagrangian \cite{Sachdevbook, Zinn}:
\beq
\label{L} L_{eff} ~=~ {{1}\over{2t}}~
\Big(\partial \theta \Big)^2 ~+~ \ldots~,
\eeq
where the dots
indicate higher derivative terms in $\theta$. The dimensionless
coefficient $1/t$ is given to leading order in the $1/N_f$
expansion by
 \beq
 \label{tcoef}
 {{1}\over{t}}~=~ N_f \, {{\rho_T}
\over {4 \pi T}} \, tanh \Big({{\rho_T} \over {2 T}}\Big)~. \eeq 

In the very low temperature limit ($T << T_c$),  
physics {\it at} momentum scales of order $T$ as well as below it
may be described by an effective theory obtained by integrating out the 
fermions and the $\rho$ fluctuations, but keeping all the Matsubara
freqauncies of $\theta$. This theory is described the  effective
chiral Lagrangian (\ref{L}), but in three dimensions rather than two.

Since the underlying theory is Abelian, this effective Lagrangian
describes a free massless scalar field $\theta$. Chiral symmetry
breaking depends on the behavior of correlation functions of
physical, and therefore single valued functions of $\theta$ 
($cos~\theta$, $sin~\theta$, or equivalently $e^{\pm i~ \theta}$).
This behavior is determined
the coefficient $t$ in the chiral Lagrangian, which is of order
$1/N_f$ or smaller unless $\rho_T / T$ is small. (This only happens
if $T$ approaches $T_c$ and we are avoiding that limit now to
insure convergence of the $1/N_f$ expansion). Analysis of this 2D
theory, including the role of vortex solutions 
\cite{Zinn}, reveals that for small $t$, it is in the
Kosterlitz-Thouless \cite{BKT} phase where the contribution from
vortices is negligible, 
and where, for example, the correlation function 
$\langle e^{i \theta(x)} \, e^{-i \theta(0)}\rangle$ 
has the characteristic power-law behavior
$\langle e^{i \theta(x)} \, e^{-i \theta(0)}\rangle \sim
x^{-t/2\pi} $~ at distances $x >> 1/T$. 
The power-law fall-off,
while corresponding to an infinite correlation length, still
indicates an absence of long range order. There is no chiral
symmetry breaking.

The transition from broken to unbroken chiral symmetry is at
$T=0$. In the limit $T \rightarrow 0 $ the parameter 
$t \rightarrow 0$, and the
range of relevance for the effective 2D theory 
and the (weakening) power-law
fall off moves off to infinity. In the zero temperature 
3D theory, the $\theta$ field develops a fixed VEV and describes 
a Goldstone boson.

At finite $T$, the analysis of the effective Abelian 2D theory
leads to the conclusion \cite{Zinn} that when $t$ becomes of
order unity, the interactions become strong, vortex solutions play
an important role, effectively renormalizing $t$, 
and a finite correlation length develops. This
is the normal phase corresponding to in-tact chiral symmetry. For
the $U(1 )$ model being discussed here, $t$ can become of
order unity only when $\rho_{T} / T \rightarrow 0$, that is, when $T
\rightarrow T_c$ from below. To be more precise, since $\rho_{T}
\sim (T_c - T) ^{1/2}$ in this limit, $t$ becomes of order unity
only when 
$(T_c - T)/T \simle 1/N_f$. 
But we have already noted that this
is the range where the $1/N_f$ expansion breaks down. The
$\rho$-field becomes light and must be included along with
$\theta$ in an effective 2D Landau-Ginzburg Lagrangian describing
the transition at $T_c$, now interpreted to be the transition from
the Kosterlitz-Thouless phase to the symmetric phase with finite
correlation length. The infrared singularities associated with the
2D description mean that the $1/N_f$ expansion is not directly
useful to describe this transition.

Finally, we comment on the behavior of the thermodynamic free
energy $F(T) = V_{eff}(\rho_T)$ for this continuous-symmetry model, and 
the proposed inequality constraint \cite{ACS} using 
the quantity $f(T) \equiv -2\pi [F(T)-F(0)] / \zeta(3)T^3$. 
In the limit $T \rightarrow \infty$,  
the model is weakly coupled (as was the
discrete-symmetry model) with the dynamics governed by a
nontrivial UV fixed point of strength $1/N_f$. To leading order in
$1/N_f$,  $f(T)$ takes the same form as in the discrete-symmetry model:
\beqa 
\label{fconT}
f(T)~=~-\, {2 N_f \over \zeta(3) T^3} \, \Big(\, {\rho_{T}^3
\over 3} - {\rho_0 \over 2}\, \rho_{T}^2 + {\rho_0^3 \over
6} ~+~ 2 T^3 \sum_{k=1}^{\infty} {(-1)^{k+1} \over k^3}\, 
\Big(1 + k {\rho_{T} \over T}\Big)\, exp[-k {\rho_{T} \over T}]\,
\Big)~. 
\eeqa
%
%\beqa \label{Vsum} V_{eff}(\rho_T)/N_f ~=~ {1 \over 3
%\pi}\,\rho_T^3 ~-~ {\rho_0\over 2 \pi}\, \rho_T^2 ~+~ {2 T^2 \over
%\pi} \Big(\rho_T \, \sum_{k=1}^{\infty} {(-1)^k \over k^2}
%e^{-\sqrt{\rho_T \over T} k} ~+~ T \sum_{k=1}^{\infty} {(-1)^k
%\over k^3} e^{-\sqrt{\rho_T \over T} k} \Big)~, \eeqa 
%
Thus to this
order,  $f_{UV} = 3\,N_f$. There will be ${\cal O}(1)$ corrections to
this result due to the weakly interacting UV fixed point.

In the low temperature limit ($T \ll T_c$), the discrete- and
continuous- symmetry models behave differently. In both cases, the
$1/N_f$ expansion for the free energy is convergent, describing in
the discrete case a mass gap and leading to $f_{IR} = 0$ to all
orders. In the continuous case, there is one massless degree of
freedom described by the $\theta$ field. 
At momentum scales  on the order of $T$ which
determine the free energy, its behaviour is fully 3D, 
governed  by an effective 3D chiral Lagrangian of the form of Eq. \ref{L}.
Since this 3D Lagrangian is non-interacting,
$f_{IR}$ may be computed exactly to give $f_{IR} = 1$ corresponding to
the single massless  degree of freedom. 
This computation of $f_{IR}$ is, in effect, an evaluation of the next to
leading term in the $V_{eff}$ given by Eq. \ref{1/N}, for $T \ll T_c$.
For this model as for the discrete symmetry model, $f_{IR} < f_{UV}$.

\section{Summary and Discussion}

We have described the finite temperature phase structure of two 3D
NJL models analyzed in a $1/N_f$ expansion where $N_f$ is the
(large) number of four-component fermions. In the more familiar case of a
discrete chiral symmetry, broken at $T = 0$, there is a transition
from broken symmetry to unbroken symmetry at a non-zero
temperature $T_c$. The low temperature phase is characterized by a
mass gap; since there are no massless degrees of freedom, the quantity 
$f_{IR}$ defined in the Section 1 takes the value $f_{IR} = 0$. 
The $1/N_{f}$ expansion breaks down as $T
\rightarrow T_c$ due to infrared singularities describable by an
effective 2D Landau-Ginzburg theory. At high temperatures, the
$1/N_{f}$ expansion is again convergent for large $N_f$
 and the theory is governed by a weak $O(1/N_{f})$ UV fixed point. 
One finds $f_{UV} = 3\, N_f$ up to corrections of $O(1/N_{f})$.

In the case of a continuous $U(1) \times U(1)$ symmetry broken at
$T = 0$, there is no spontaneous breaking of the chiral symmetry
at $T \neq 0$, although a phase transition still exists at a
non-zero temperature $T_c$. The normal high-temperature symmetric
phase changes at $T_c$ to a low-temperature 
phase with its  low momentum components ($p \ll T$) described by 
an effective 2D
theory in the Kosterlitz-Thouless (K-T) phase and with a power-law
behaviour of correlation functions for $x \gg 1/T$. 
In both finite-$T$ phases, the continuous chiral
symmetry is unbroken. The transition from the K-T phase
to the broken phase is at $T = 0$. There is a single massless
degree of freedom for $T < T_c$ which becomes a conventional 
Goldstone boson in the limit $T \rightarrow 0$. 
As in the case of discrete symmetry, the $1/N_f$ expansion breaks 
down at the phase transition $ T \sim T_c$ due to 
2D infrared singularities, but is well behaved away
from the transition for large $N_f$.
We have nothing to say in this paper about the phase 
structure of NJL models at small $N_f$ \cite{EB}.
 
The quantity $f_{IR}$ 
defined in the Section 1 takes the value $f_{IR} = 1$,
reflecting the presence of a single massless 
degree of  freedom below $T_c$. 
The UV fixed point leads to the value for $f_{UV} = 3\, N_f$ 
up to corrections of $O(1/N_{f})$ as in the discrete case.

For both models, the exploration of the transition at $T = T_c$
requires methods that are non-perturbative in $1/N_{f}$. For the
discrete case, $\epsilon$-expansion methods 
($4 - \epsilon$ dimensions)
may be brought to bear
although convergence is problematic since $\epsilon = 2$ . In the
continuous case, the conventional $\epsilon$ expansion is not useful since the
model is in the broken phase for $T < T_c$ for any $\epsilon < 2$. 
In both cases, the
breakdown of the $1/N_{f}$ expansion  is associated with
finite-temperature infrared divergences characteristic of a theory
in less than four space-time dimensions, and has no direct
counterpart in four dimensional theories.

It is interesting to generalize our discussion to NJL models with
larger continuous chiral symmetries. Suppose, for example, that
the continuous $U(1) \otimes U(1)$ symmetry of (\ref{njlcon}) is
extended to a $U(n) \otimes U(n)$ symmetry which at zero
temperature breaks to the diagonal $U(n)$ producing $n^2$
Goldstone modes. 
For the case $n^2 << N_f$, the $1/N_f$
expansion may be used to analyze the symmetry breaking pattern at zero and
finite temperatures. An order by  order analysis concludes that for 
$ T > T_{c} \sim \rho_0$ (where $\rho_0$ is a zero temperature vev of an
auxilliary field) the model will be in the symmetric phase. 
As in the abelian case the $1/N_f$ expansion breaks down when $T \sim T_c$
due to 2D infrared singuliarities.
For $T < T_c$, the fermions are massive, as in the abelian model, 
and may be integrated out to describe 
lower energy physics. At momentum scales below $T$,  an
effective 2D chiral Lagrangian emerges, which naturally breaks into
two parts. One is the free abelian model we considered previously and the
other is an $SU(n) \times SU(n)$ chiral Lagrangian. The latter is
interacting and asymptotically free \cite{Sachdevbook, Pol, PS}. 

The
abelian model will be in the KT phase as long as $T$ is not close to
$T_c$, ensuring that the coupling strength is small -- of order $1/N_f$. 
At intermediate momentum scales, $p < T$, the coupling strength of the 
non-abelian part is also small -- of order $1/N_f$.  
The $\beta$-function for this coupling $\tilde{t}$ is 
$\beta_{\tilde{t}} \sim - b \, \tilde{t}\,^2$ where the constant 
$b$ is positive and of  order $n^2/(4\pi)$ for $ n > 1$. 
Thus the effective $\tilde{t}$ coupling runs, becoming of order
unity at ultra-low scales 
$\mu_{IR} \sim T ~exp(- 2 \pi \cdot N_f/n^2 \cdot \rho_T/T)$. 
As in the case of the $U(1) \otimes U(1)$ model, unit
strength is expected to disorder the system leading to a finite
correlation length,  now of order $1 /\mu_{IR}$.  But since this happens for
any finite $T < T_c$, the non-abelian theory is in the ordinary
symmetric phase, not the KT phase\footnote{This conclusion is modified in 
the presence of the Wess-Zumino-Witten term. However, because the underlying
3D fermionic theory does not carry anomalies, the effective low energy
scalar theory is  also anomaly free and no WZW term is generated 
\cite{Fendley}.}.
For the abelian part only, there is a KT
transition as $T \rightarrow T_c$.  
The value of the $f_{IR}$ is $n^2$
because although the light scalars of the non-abelian part of the effective
theory are massive, their mass $m_{SU(N)} \sim \mu_{IR}$ 
vanishes exponentially  as $T \rightarrow 0$. Thus they are effectively
massless in this limit.
Clearly, for generic 
$n^2 \ll N_f$ the inequality $f_{IR} \leq f_{UV}$ will always be 
satisfied since $f_{UV} = 3\, N_f$.

If the $U(n) \otimes U(n)$ model is arranged
so that $n^2$ becomes of order $N_f$, the $1/N_f$ expansion breaks
down at all $T$ due to the large number of scalar degrees of
freedom (${\cal O}(n^2)$) that are formed. 
%
%The scale $\mu_{IR}$ is of order $T$ unless $T << \rho_T$. 
%
It is beyond the scope of this paper to analyze the
model when $n^2 \sim N_f$, but we note that it may well be accessible to
a combined $1/N_f$, $1/n^2$ expansion. In the limit $ T <<
\rho_T$, the model is described by an effective, 3D chiral
Lagrangian at the scales of $T$, and the
$O(N)$ models of this sort have been analyzed
using the $1/N$ expansion \cite{Sachdev-Zinn}. 

\bigskip

\noindent{\bf  Acknowledgements}
\bigskip

We are grateful for several helpful discussions with Nick Read, Subir
Sachdev and L.C.R. Wijewardhana.  
We also thank Nick Read for a careful reading of the manuscript.

\end{document}